\let\orilabel\label
\documentclass[aps,pra,preprint,a4paper,showpacs,showkeys,superscriptaddress,nofootinbib]{revtex4-1}
\let\label\orilabel

\usepackage{latexsym}
\usepackage{amsmath,amssymb}
\usepackage{graphicx}
\usepackage{subfigure}
\usepackage{xcolor}
\usepackage{physics}
\usepackage{cancel}
\usepackage{tikz}
\usepackage{multirow,tabularx}

\newcolumntype{C}[1]{>{\centering\arraybackslash}p{#1}}
\usepackage{hyperref}     
\hypersetup{colorlinks,%
  citecolor=blue,%
  linkcolor=cyan,%
}
\usepackage[titletoc]{appendix}
\usepackage{enumerate}
\graphicspath{{./Figures/}}

\begin{document}


\title{Temperatures of AdS\texorpdfstring{$_2$}{2} black holes and holography revisited}

\author{Wontae Kim}%
\email[]{wtkim@sogang.ac.kr}%
\affiliation{Department of Physics, Sogang University, Seoul, 04107,
	Republic of Korea}%
\affiliation{Center for Quantum Spacetime, Sogang University, Seoul 04107, Republic of Korea}%

\author{Mungon Nam}%
\email[]{clrchr0909@sogang.ac.kr}%
\affiliation{Department of Physics, Sogang University, Seoul, 04107,
	Republic of Korea}%
\affiliation{Center for Quantum Spacetime, Sogang University, Seoul 04107, Republic of Korea}%
\date{\today}

\begin{abstract}
In a dilaton gravity model, we revisit the calculation of the temperature of an evaporating black hole that is initially formed by a shock wave, taking into account the quantum backreaction. Based on the holographic principle, along with the assumption of a boundary equation of motion, we show that the black hole energy is maintained for a while during the early stage of evaporation. Gradually, it decreases as time goes on and eventually vanishes. Thus, the Stefan-Boltzmann law tells us that the black hole temperature, defined by the emission rate of the black hole energy, starts from zero temperature and reaches a maximum value at a critical time, and finally vanishes. It is also shown that the maximum temperature of the evaporating black hole never exceeds the Hawking temperature of the eternal AdS$_{2}$ black hole. We discuss physical implications of the initial zero temperature of the evaporating black hole.


\end{abstract}
%



\maketitle


\raggedbottom

\section{Introduction}
\label{sec:introduction}
Many years ago, Jackiw~\cite{Jackiw:1984je} and Teitelboim~\cite{Teitelboim:1983ux} proposed a dilaton gravity model for two-dimensional anti-de Sitter (AdS$_2$) space.
It is known that any excitation with finite energy causes significant backreaction on the AdS spacetime~\cite{Maldacena:1998uz,Strominger:1998yg}.
Almheiri and Polchinski (AP)~\cite{Almheiri:2014cka} further developed a modified model that provides a regulated quantum backreaction, which allows one to set up a more meaningful holographic dictionary through the analysis of the black hole in the Hartle-Hawking state~\cite{Hartle:1976tp,Israel:1976ur}.
From the fact that the time coordinate becomes dynamical on the boundary~\cite{Maldacena:2016upp}, the black hole in the Unruh state~\cite{Unruh:1976db} was investigated by assuming a dynamical boundary equation of motion in the AP model~\cite{Engelsoy:2016xyb}. In fact, in a wide variety of cases of two-dimensional toy models, there has been much attention to the backreaction issue that exhibits black hole formation and evaporation~\cite{Callan:1992rs,Giddings:1992ff,Russo:1992ax,Keski-Vakkuri:1993ybv,Liberati:1994za,Navarro-Salas:1995lmi,Fabbri:1995bz,Bose:1995bk,Nojiri:1997sr,Kim:1998wy,Spradlin:1999bn,Fabbri:2005mw,Almheiri:2019psf}.

In particular, the black hole temperature characterizing Hawking radiation plays an important role in black hole thermodynamics, along with the black hole entropy. The common wisdom is that black holes in thermal equilibrium have a constant temperature that can be calculated from the surface gravity method on the black hole side or from the Stefan-Boltzmann law for radiation in a bath side. On the other hand, in evaporating black holes, the calculation of the black hole temperature would be non-trivial since the quasi-static temperature is also assumed to be time-dependent, and thus sensitive to the backreaction of geometry. A recent study in the AP model~\cite{Engelsoy:2016xyb} shows that the initial temperature of the black hole starts from the Hawking temperature and finally vanishes when taking into account the quantum backreaction. In the end, the black hole temperature approaches zero, which implies that the radiation temperature should also be zero in the quasi-equilibrium assumption. Now, one might wonder how the black hole temperature could suddenly become the Hawking temperature as soon as the black hole forms from collapsing matter. It might take some time for the black hole to be thermalized and become a thermal object~\cite{Hayden:2007cs}.

In this work, we revisit the calculations of the temperature of the evaporating AdS$_2$ black hole in the AP model. We adopt the dynamical equation of motion proposed in Ref.~\cite{Engelsoy:2016xyb} and properly take into account the effect of quantum backreaction. 
In particular, we obtain the vacuum solution \eqref{eq:bdy Omega in evap} from the Unruh boundary conditions, which is different from the previous result in Ref.~\cite{Engelsoy:2016xyb}.
Consequently, we show that the black hole temperature is initially zero and reaches a maximum temperature less than the Hawking temperature, before finally approaching zero temperature.

The organization of the paper is as follows: In Sec.~\ref{sec:Preliminaries}, we employ holographic renormalization process to derive the holographic stress tensor in the AP model and present the general form of the black hole energy associated with various vacuum conditions. In Sec.~\ref{sec:The Black hole in Equilibrium},
we compute the energy and temperature of the eternal AdS$_2$ black hole by utilizing the energy expression presented in Sec.~\ref{sec:Preliminaries}.
Our calculations confirm the well-established outcomes in the Hartle-Hawking vacuum state.
In Sec.~\ref{sec:The Black hole in Evaporation}, we study the black hole energy in the evaporating AdS$_2$ black hole and find a different type of temperature in the Unruh vacuum state. Finally, we give our conclusions and discuss our results in Sec.~\ref{sec:conclusion}.

\section{The Holographic stress tensor}
\label{sec:Preliminaries}
In this section, we introduce the two-dimensional dilaton gravity model, which is described by the action consisting of the AP action with classical and quantum matter \cite{Almheiri:2014cka}:
\begin{align}
	S &= S_{\rm AP}+ S_{\rm matt}= S_{\rm AP} +S_{\rm cl} + S_{\rm qt},\label{eq:total action}\\
	S_{\rm AP} &= \frac{1}{16\pi G}\int\dd[2]{x}\sqrt{-g}\left[ \Phi^2(R+2)-2 \right] + \frac{1}{8\pi G}\int \dd x^0 \sqrt{-\gamma}\Phi^2 K,\label{eq:AP action}\\
	S_{\rm cl} &= -\frac{1}{2}\int\dd[2]{x}\sqrt{-g}\sum_{i=1}^{N}(\nabla f_i)^2,\label{eq:classical matter action}\\
	S_{\rm qt} &= -\frac{N}{24\pi}\int\dd[2]{x}\sqrt{-g}\left[ (\nabla \chi)^2 + \chi R \right] - \frac{N}{12\pi}\int\dd x^0\sqrt{-\gamma}\chi K.\label{eq:quantum matter action}
\end{align}
Here, $\Phi^2$, $f_i$, and $K$ represent the dilaton field, scalar matter fields, and the extrinsic curvature, respectively.
The auxiliary scalar field $\chi$ is used to localize the non-local Polyakov action \cite{Polyakov:1981rd}. (For a review, see \cite{Vassilevich:2003xt}.)
The matter contribution is usually subdominant compared to the gravitational contribution.
This observation is based on the matter CFT calculation~\cite{Yang:2018gdb,Mertens:2022irh}.
The one-loop Polyakov action for matter fields will be considered within the large $N$ approximation with $\Phi^{-2}N$ being fixed
so that the matter part becomes comparable to the gravitational part~\cite{Strominger:1994tn}.

In the conformal gauge of $\dd s^2 = -e^{2\omega(u,v)}\dd u\dd v$, where $u = t + \sigma$ and $v = t - \sigma$, the equations of motion are given as follows:
\begin{align}
	4\partial_u\partial_v \omega + e^{2\omega} &= 0,\label{eq:metric eq}\\
	2\partial_u\partial_v \Phi^2 + (\Phi^2-1)e^{2\omega} &= 16\pi G T_{uv},\label{eq:dilaton uv}\\
	-e^{2\omega}\partial_{u}(e^{-2\omega}\partial_u\Phi^2) &= 8\pi G T_{uu},\label{eq:dilaton uu}\\
	-e^{2\omega}\partial_{v}(e^{-2\omega}\partial_v\Phi^2) &= 8\pi G T_{uu},\label{eq:dilaton vv}\\
	\partial_u\partial_vf_i &= 0,\label{eq:classical field eq}\\
	\partial_u\partial_v(\chi + \omega) &= 0,\label{eq:aux field eq}
\end{align}
where $T_{\mu\nu}= -\frac{2}{\sqrt{-g}}\fdv{S_{\rm matt}}{g^{\mu\nu}} = T_{\mu\nu}^{\rm cl} + T_{\mu\nu}^{\rm qt}$.
The stress tensors for quantum matter are expressed as
\begin{align}
	T_{uu}^{\rm qt} &= \frac{N}{12\pi}\left[ (\partial_{u}\chi)^2 - \partial_{u}^2\chi + 2\partial_{u}\omega\partial_{u}\chi \right], \label{eq:qt stress tensor uu}\\
	T_{vv}^{\rm qt} &= \frac{N}{12\pi}\left[ (\partial_{v}\chi)^2 - \partial_{v}^2\chi + 2\partial_{v}\omega\partial_{v}\chi \right], \label{eq:qt stress tensor vv}
\end{align}
where $T_{uv}^{\rm qt} = \frac{N}{12\pi}\partial_u\partial_v\chi$.

The general solution to the metric equation \eqref{eq:metric eq} is given by the AdS$_2$ spacetime:
\begin{equation}
	\label{eq:gen sol metric}
	\dd s^2 = -\frac{4}{(x^+ - x^-)^2}\dd x^+\dd x^-=-\frac{4\partial_u x^+(u) \partial_v x^-(v)}{(x^+(u) - x^-(v))^2}\dd u \dd v,
\end{equation}
where $x^{\pm} = x^0 \pm x^1$ are general monotonic functions of the light-cone coordinates
$x^+ = x^+(u)$ and $x^- = x^-(v)$,
 respectively.
The boundary of AdS$_2$ is located at $x^+ = x^-$.
From Eq.~\eqref{eq:dilaton uv} with the constraint equations \eqref{eq:dilaton uu} and \eqref{eq:dilaton vv}, the dilaton field can be solved as \cite{Almheiri:2014cka,Engelsoy:2016xyb}
\begin{equation}
	\label{eq:gen sol dilaton}
	\Phi^2 = 1 + \frac{GN}{3} + \frac{a}{x^+ - x^-}\left[ 1 - \kappa(I_+ + I_-)\right],
\end{equation}
where $ a $ is an integration constant and $\kappa = 8\pi G/a$. In Eq.~\eqref{eq:gen sol dilaton}, the sources are $I_+ = \int^{\infty}_{x^+}\dd s(s - x^+)(s-x^-)T_{++}^{\rm cl}(s) + \int^{\infty}_{x^+}\dd s(s - x^+)(s-x^-)T_{++}^{\rm qt}(s) = I_+^{\rm cl} + I_+^{\rm qt}$ and $I_- = \int^{x^-}_{-\infty}\dd s(s-x^+)(s-x^-) T_{--}^{\rm cl}(s) + \int^{x^-}_{-\infty}\dd s(s-x^+)(s-x^-) T_{--}^{\rm qt}(s) = I_-^{\rm cl} + I_-^{\rm qt}$.
Note that the choice of $ a>0 $ prevents the strong coupling singularity at which $ \Phi^2 = 0 $ from reaching the boundary within a finite amount of time. In Eq.~\eqref{eq:classical field eq}, the infalling classical matter field is solved as $f_i (u,v) =f_i (u)+f_i(v)$. Finally, the auxiliary field $ \chi $ in Eq.~\eqref{eq:aux field eq} is also obtained as
\begin{equation}
	\label{eq:gen sol auxiliary}
	\chi(u,v) = -\omega(u,v) + \Omega_u(u) + \Omega_v(v),
\end{equation}
where $ \Omega_{u}(u),\Omega_{v}(v) $ are responsible for quantum vacuum states which play an important role in our calculations.

Plugging Eqs.~\eqref{eq:gen sol metric} and \eqref{eq:gen sol auxiliary} into Eqs.~\eqref{eq:qt stress tensor uu} and \eqref{eq:qt stress tensor vv}, we can express quantum matter as
	\begin{equation}
		\label{eq:qt stress tensor sol}
		T_{uu}^{\rm qt}(u) = \frac{N}{24\pi}\{ x^+(u),u \} + :T^{\rm qt}_{uu}: ,\quad T_{vv}^{\rm qt}(v) = \frac{N}{24\pi}\{ x^-(v),v \} + :T^{\rm qt}_{vv}:
	\end{equation}
	where the Schwarzian derivative is defined as $\{ F(x),x \} = \frac{\partial^3_x{F}(x)}{\partial_x{F}(x)} - \frac{3}{2}\left( \frac{\partial^2_x{F}(x)}{\partial_x{F}(x)} \right)^2$ and the normal ordered stress tensors are
	\begin{equation}
		\label{eq:omega}
		:T^{\rm qt}_{uu}: = \frac{N}{12\pi}\left[ (\partial_u\Omega_u)^2 - \partial^2_u\Omega_u \right],\quad :T^{\rm qt}_{vv}: = \frac{N}{12\pi}\left[ (\partial_v\Omega_v)^2 - \partial^2_v\Omega_v \right].
	\end{equation}
The normal ordered stress tensors are measured by local observers linked to a particular coordinate system. They are not true tensors since they break general covariance.

For a dynamical boundary time defined by $x^+(t) = x^-(t) = \tau(t)$ at the AdS boundary, one can consider the coordinates for a slightly perturbed boundary with a regulator $\epsilon$ as $x^0(t,\epsilon) = \frac{1}{2}[x^+(t+\epsilon)+x^-(t-\epsilon)] = \tau(t)$ and $x^1(t,\epsilon) = \frac{1}{2}[x^+(t+\epsilon) - x^-(t-\epsilon)] = \epsilon\dot{\tau}(t)$,
where $\epsilon$ is a constant and $\epsilon\dot{\tau}(t)$ represents the distance between the dynamical boundary from the unperturbed AdS$_2$ boundary. We assume, as proposed in Ref.~\cite{Engelsoy:2016xyb}, that the dilaton has the same asymptotic form near the boundary as in the Poincar\'e patch:
\begin{equation}
	\label{eq:asymptote dilaton cond}
	\Phi^2(t) = 1 + \frac{GN}{3} + \frac{a}{2\epsilon\dot{\tau}(t)}\left[ 1 - \kappa(I_+(t) + I_-(t)) \right] = \frac{a}{2\epsilon},
\end{equation}
which implies that $\dot{\tau}(t) = 1 - \kappa[I_+(t) +I_-(t)]$, where $I_+(t) = \int_{\tau}^{\infty}\dd s (s-\tau)^2T_{++}(s)$ and $I_-(t) = \int_{-\infty}^{\tau}\dd s(s-\tau)^2T_{--}(s)$. Thus, the equation of motion for the dynamical boundary time reads
\begin{equation}
	\label{eq:bdy equation}
	\frac{1}{2\kappa}\dv[2]{t}\log \dot{\tau} + (P_+(t) - P_-(t))\dot{\tau} = 0,
\end{equation}
where $P_+ = \int_{\tau}^{\infty}\dd s\, T_{++}(s)$ and $P_- = -\int_{-\infty}^{\tau}\dd s\, T_{--}(s)$. Interestingly, one more differentiation of Eq.~\eqref{eq:bdy equation} with respect to $t$ leads to a very useful relation~\cite{Mertens:2022irh}:
\begin{equation}
	\label{eq:bdy eom}
	-\frac{1}{2\kappa}\dv{t}\{ \tau,t \} = T_{vv}(t) - T_{uu}(t) = :T^{\rm qt}_{vv}(t): - :T^{\rm qt}_{uu}(t):.
\end{equation}
This result was also derived from the Hamiltonian formulation
by using the holographic energy written as the Schwarzian derivative~\cite{Engelsoy:2016xyb}.

In Eqs.~\eqref{eq:gen sol metric}, \eqref{eq:gen sol dilaton}, and \eqref{eq:gen sol auxiliary}, asymptotic expansions of fields near the boundary are calculated as
\begin{align}
	e^{2\omega(t,\epsilon)} &= \frac{1}{\epsilon^2} + \frac{2}{3}\{ \tau,t \} + \mathcal{O}(\epsilon^2),\label{eq:bdy exp metric}\\
	\Phi^2 (t,\epsilon) &= \frac{a}{2\epsilon} + 1 + \frac{GN}{3} - \frac{a}{3}\{ \tau,t \}\epsilon + \mathcal{O}(\epsilon^2),\label{eq:bdy exp phi}\\
	\chi(t,\epsilon) &= \log \epsilon + \Omega_u(t) + \Omega_v(t) + \epsilon\dot{\Omega}_u(t) - \epsilon\dot{\Omega}_v(t)+\mathcal{O}(\epsilon^2).\label{eq:bdy exp chi}
\end{align}
Next, the holographic renormalized action can be constructed as $S_{\rm R} = S + S_{\rm ct}$ adding the counter term
$S_{\rm ct} = \frac{1}{8\pi G}\int\dd t\sqrt{-\gamma}(1-\Phi^2) - \frac{N}{24\pi}\int\dd t\sqrt{-\gamma}$.
Thus, the boundary stress tensor in dual field theory can be obtained by varying the on-shell bulk action with respect to the boundary metric \cite{Balasubramanian:1999re}: $\langle \hat{T}_{tt} \rangle = -\frac{2}{\sqrt{-\gamma}}\fdv{S_{\rm R}}{\hat{\gamma}^{tt}} = \lim_{\epsilon\to 0}\frac{-2\epsilon}{\sqrt{-\gamma(\epsilon)}}\fdv{S_{\rm R}}{\gamma^{tt}(\epsilon)}$,
where the boundary metric $\hat{\gamma}_{tt}$ is related to the induced metric $\gamma_{tt}$ through $\hat{\gamma}_{tt} = \lim\limits_{\epsilon\to 0}\epsilon^2\gamma_{tt}(\epsilon)$. Hence, the boundary stress tensor can be obtained as \cite{Almheiri:2014cka}
\begin{equation}
	\label{eq:bdy stress tensor}
	\langle \hat{T}_{tt} \rangle = \frac{\epsilon}{8\pi G}\left( e^{\omega}\partial_{\epsilon}\Phi^2 -e^{2\omega}(1-\Phi^2) \right) - \frac{N\epsilon}{24\pi}\left( 2e^{\omega}\partial_{\epsilon}\chi -e^{2\omega} \right).
\end{equation}
Plugging the asymptotic forms of the fields \eqref{eq:bdy exp metric}, \eqref{eq:bdy exp phi}, and \eqref{eq:bdy exp chi} into Eq.~\eqref{eq:bdy stress tensor}, one can obtain a final expression for the boundary stress tensor as \cite{Engelsoy1117777}
\begin{equation}
	\label{eq:bdy stress tensor wrt tau}
	\langle \hat{T}_{tt} \rangle = E(t) = -\frac{1}{2\kappa}\{ \tau,t \} - \frac{N}{12\pi}(\dot{\Omega}_u(t) - \dot{\Omega}_v(t)).
\end{equation}
In fact, $\Omega_u$ and $\Omega_v$ will be determined by choosing quantum vacuum states of the normal ordered stress tensors \eqref{eq:omega}. In the subsequent sections, we shall choose $\Omega_u$ and $\Omega_v$ for an eternal black hole and an evaporating black hole, respectively.

\section{The eternal black hole}
\label{sec:The Black hole in Equilibrium}
In the eternal AdS$_2$ black hole, we investigate the energy expression~\eqref{eq:bdy stress tensor wrt tau} and obtain the black hole temperature from normal ordered stress tensors in the Hartle-Hawking vacuum state instead of the surface gravity method. The eternal black hole can be obtained by considering the vanishing classical stress tensors as $T_{--}^{\rm cl} = T_{++}^{\rm cl} = T_{+-}^{\rm cl} = 0$,
and thus, the dilaton field \eqref{eq:gen sol dilaton} takes the following form:
\begin{equation}
	\label{eq:dilaton sol eternal}
	\Phi^2 = 1 + \frac{GN}{3} + \frac{a}{x^+ - x^-}\left[ 1 - \kappa E_0 - \kappa(I_+^{\rm qt} + I_-^{\rm qt}) \right]
\end{equation}
without modification of the metric \eqref{eq:gen sol metric}, where $E_0$ is an integration constant.
In thermal equilibrium, the appropriate condition, called the Hartle-Hawking vacuum state, is given by \cite{Almheiri:2014cka,Spradlin:1999bn,Pedraza:2021cvx}
\begin{equation}
	\label{eq:equilibrium cond in general coord}
	:T^{\rm qt}_{\pm\pm}(x^{\pm}): = 0.
\end{equation}
The vacuum condition \eqref{eq:equilibrium cond in general coord} can be equivalently rewritten in terms of $(u,v)$ coordinates as
\begin{equation}
	\label{eq:equilibrium cond in uv coord}
	:T^{\rm qt}_{uu}(u): = -\frac{N}{24\pi}\{ x^+(u), u \},\quad :T^{\rm qt}_{vv}(v): = -\frac{N}{24\pi}\{ x^-(v), v \}
\end{equation}
under the anomalous transformation of Eq.~\eqref{eq:equilibrium cond in general coord}.
Thus, in this boundary condition, Eq.~\eqref{eq:omega} takes the following form: $(\partial_{u}\Omega_u)^2 - \partial^2_u\Omega_u = -\frac{1}{2}\{ x^+(u),u \}$ and $
	(\partial_{v}\Omega_v)^2 - \partial^2_v\Omega_v = -\frac{1}{2}\{ x^-(v),v \}$,
which can be solved as
\begin{equation}
	\label{eq:Omega diff sol w/ b}
	\Omega_u(u) = \frac{1}{2}\log\frac{\partial_ux^+(u)}{\left( 1 + b^+x^+(u) \right)^2}+b^+_0,\quad \Omega_v(v) = \frac{1}{2}\log\frac{\partial_vx^-(v)}{\left( 1 + b^-x^-(v) \right)^2 }+b^-_0,
\end{equation}
where $b^{\pm}$ and $b^{\pm}_0$ are integration constants, but the latter constants are set to zero without loss of generality.

In the Hartle-Hawking vacuum state~\eqref{eq:equilibrium cond in general coord}, the dilaton field \eqref{eq:dilaton sol eternal} takes the form:
\begin{equation}
	\label{eq:dilaton sol HH}
	\Phi^2(u,v) = 1 + \frac{GN}{3} + \frac{a}{x^+ - x^-}\left[ 1 - \kappa E_0x^+x^- \right].
\end{equation}
The metric \eqref{eq:gen sol metric} remains unaffected by the matter.
Then, the coordinate transformations
\begin{equation}
	\label{eq:static coord}
	x^+(u) = \frac{1}{\sqrt{\kappa E_0}}\tanh(\sqrt{\kappa E_0} u),\quad x^-(v) = \frac{1}{\sqrt{\kappa E_0}}\tanh(\sqrt{\kappa E_0} v)
\end{equation}
render the metric and the dilaton field static:
\begin{align}
	e^{2\omega}(u,v) &= \frac{4\kappa E_0}{\sinh^2\sqrt{\kappa E_0}(u-v)},\label{eq:metric sol in equilibrium}\\
	\Phi^2(u,v) &= 1 + \frac{GN}{3} + a\sqrt{\kappa E_0}\coth\sqrt{\kappa E_0}(u-v).\label{eq:dilaton sol in equilibrium}
\end{align}
Next, the regularity condition on the horizon of the static auxiliary field $\chi$ in Eq.~\eqref{eq:gen sol auxiliary} is given as
\begin{equation}
\chi(u,v) = \log(\frac{\sinh\sqrt{\kappa E_0}(u-v)}{2\sqrt{\kappa E_0}}) - \sqrt{\kappa E_0}(u-v)\label{eq:chi sol in equilibrium}
\end{equation}
by requiring that
\begin{equation}
	\label{eq:Omega sol in equilibrium}
	\Omega_u(u) = -\sqrt{\kappa E_0} u,\quad \Omega_v(v) = \sqrt{\kappa E_0}v.
\end{equation}

At the AdS$_2$ boundary, the dynamical boundary time obtained from Eq.~\eqref{eq:static coord} is given by

\begin{equation}
	\label{eq:tau in equilibrium}
	\tau(t) = \frac{1}{\sqrt{\kappa E_0}}\tanh\sqrt{\kappa E_0}t,
\end{equation}
and Eq.~\eqref{eq:Omega sol in equilibrium} also becomes
\begin{equation}
	\label{eq:bdy Omega in equilibrium}
	\Omega_u(t) = -\sqrt{\kappa E_0}t, \quad \Omega_v(t) = \sqrt{\kappa E_0}t.
\end{equation}
By substituting Eqs.~\eqref{eq:tau in equilibrium} and \eqref{eq:bdy Omega in equilibrium} into Eq.~\eqref{eq:bdy stress tensor wrt tau}, we obtain the boundary stress tensor as
\begin{equation}
	\label{eq:bdy stress tensor in equilibrium}
	E= E_0 + \frac{N}{6\pi}\sqrt{\kappa E_0}.
\end{equation}
For $a=1$, the black hole energy turns out to be the same as the result in Ref.~\cite{Almheiri:2014cka}.
The relation \eqref{eq:bdy eom} is trivial in that $\{ \tau,t \}=-2\kappa E_0$, which is just a constant in the Hartle-Hawking vacuum state.
In addition, the black hole temperature identified with the radiation temperature in thermal equilibrium can be given by the Stefan-Boltzmann law of $\sigma T^2=:T^{\rm qt}_{uu}(t):+ :T^{\rm qt}_{vv}(t):$ \cite{Landsberg:1989,Fabbri:2005mw}, where the Stefan-Boltzmann constant is $\sigma =N(\pi/6)$.
In the boundary limit, Eq.~\eqref{eq:equilibrium cond in uv coord} reduces to $:T^{\rm qt}_{uu}(t): = -\frac{N}{24\pi}\{ \tau(t), t \}$ and $:T^{\rm qt}_{vv}(t): = -\frac{N}{24\pi}\{ \tau(t), t \}$ so that the black hole temperature can be obtained as $T=\frac{1}{\pi}\sqrt{\kappa E_0}=T_{\rm H}$.
As expected, the temperature is the same as the result derived from the surface gravity method using the black hole metric.

\section{The evaporating black hole}
\label{sec:The Black hole in Evaporation}
In this section, we consider the evaporating black hole formed by an infalling shock wave and study the time-dependent black hole energy and temperature. Let us choose the classical matter of an infalling pulse of energy $ E_0 $ described by $T_{--}^{\rm cl} = E_0\delta(x^-)$ and $ T_{++}^{\rm cl} = T_{+-}^{\rm cl} = 0$.
Then, the dilaton field becomes time-dependent as
\begin{equation}
	\label{eq:dilaton sol evap}
	\Phi^2 = 1 + \frac{GN}{3} + \frac{a}{x^+ - x^-}\left[ 1 - \kappa E_0\theta(x^-)x^+x^- - \kappa(I_+^{\rm qt} + I_-^{\rm qt}) \right]
\end{equation}
with the metric solution \eqref{eq:gen sol metric}.
Assuming the perfect absorption condition at the boundary and no outgoing matter on the future horizon,
we obtain the following vacuum conditions:
\begin{equation}
	\label{def:evap cond}
	:T^{\rm qt}_{++}: = 0,\quad :T_{vv}^{\rm qt}(t): = 0,
\end{equation}
where the condition $:T^{\rm qt}_{++}: = 0$ means $:T_{uu}^{\rm qt}(u): = -\frac{N}{24\pi}\{ x^+,u \} $.
Now, Eq.~\eqref{eq:omega} compatible with the boundary condition \eqref{def:evap cond} leads to the following solutions
\begin{equation}
	\label{eq:Omega sol in evap}
	\Omega_u(u) = \frac{1}{2}\log\frac{\partial_{u}x^+(u)}{(1+b^+x^+(u))^2}+b^+_0,\quad \Omega_v(t) = -\log(1+b^-t)+b^-_0,
\end{equation}
where $b^\pm$ and $b^\pm_0$ are integration constants and $b^\pm_0 =0$ without loss of generality.
At the boundary, we have
\begin{equation}
	\label{eq:bdy Omega in evap}
	\Omega_u(t) = \frac{1}{2}\log\frac{\dot{\tau}(t)}{(1+b^+\tau(t))^2},\quad \Omega_v(t) = -\log(1+b^-t).
\end{equation}
Note that these solutions are different from the previous result in Ref.~\cite{Engelsoy:2016xyb}.
Inserting Eq.~\eqref{eq:bdy Omega in evap} into Eq.~\eqref{eq:bdy stress tensor wrt tau}, we can obtain the boundary stress tensor as
\begin{equation}
	\label{eq:bdy stress tensor unruh}
	E(t) = -\frac{1}{2\kappa}\{ \tau,t \} - \frac{N}{12\pi}\left( \frac{\ddot{\tau}}{2\dot{\tau}} - \frac{b^+\dot{\tau}}{1+b^+ \tau}\right) - \frac{N}{12\pi} \left(\frac{b^-}{1+b^- t} \right),
\end{equation}
where the constants $b^+$ and $b^-$ will be fixed by physical requirements later.

\begin{figure}
	\begin{center}
		\subfigure[]{\includegraphics[width=.45\textwidth]{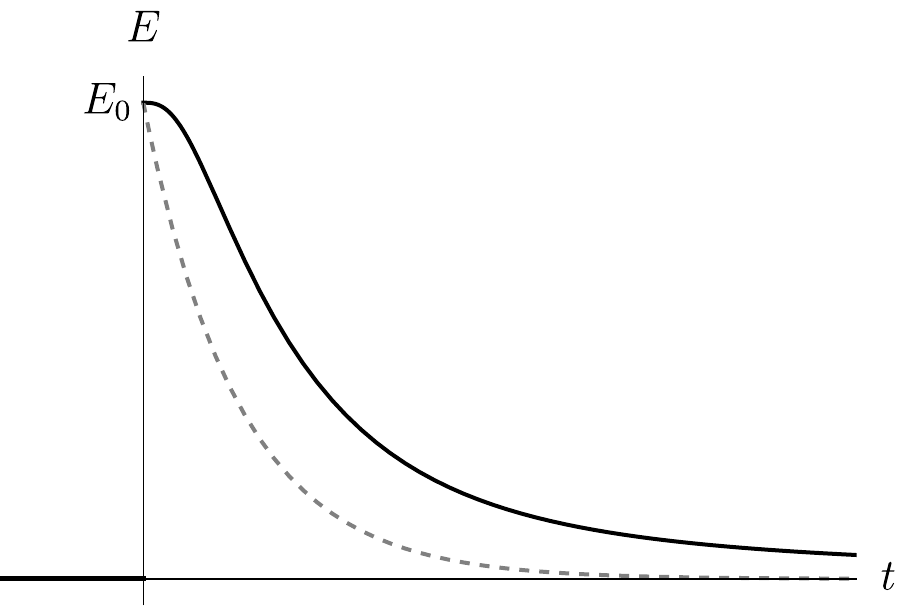}\label{fig:energy}}
		\quad
		\subfigure[]{\includegraphics[width=.45\textwidth]{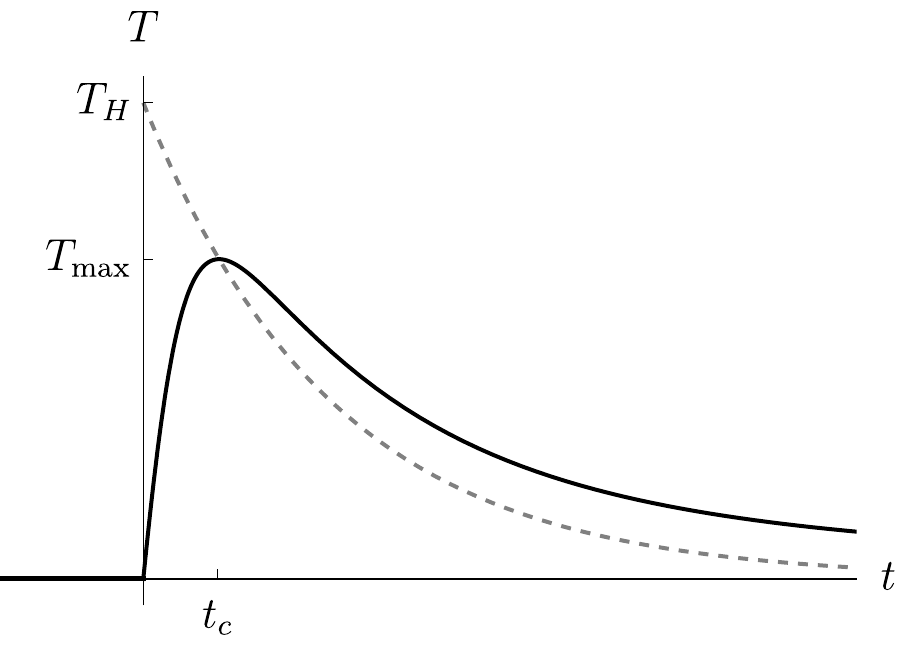}\label{fig:temperature}}
	\end{center}
\caption{For $A = 0.5$, and $\kappa E_0 = 1$, the solid curves represent the black hole energy (a) and the black hole temperature (b) under
the vacuum conditions $\Omega_u(t) = (1/2)\log \dot{\tau}(t)$ and $\Omega_v(t) = 0$, while the dashed curves represent the results
when $\Omega_u - \Omega_v$ is constant.
In Fig.~(a), the black hole energy is discontinuous at $t=0$ due to the shock wave.
In Fig.~(b),
the black hole temperature described by the solid curve is continuous at $t=0$ and the maximum temperature
occurs at a critical time $t_c$.}
\end{figure}

On the other hand, the relation~\eqref{eq:bdy eom} in the vacuum condition \eqref{def:evap cond} can be simplified as follows:
\begin{equation}
	\label{eq:bdy eom in evap}
	-\frac{1}{2\kappa} \dv{t}\{ \tau, t \} = \frac{N}{24\pi}\{ \tau,t \}
\end{equation}
which was actually obtained by using
the time evolution of the dynamical boundary equation~\eqref{eq:bdy equation}
and the boundary condition~\eqref{def:evap cond}
without resort to the holographic stress tensor~\eqref{eq:bdy stress tensor unruh}.
At $t=0$, we require the conditions:
\begin{equation}\label{eq:gluing cond}
  \tau(0) = 0,\quad \dot{\tau}(0) = 1,\quad \ddot{\tau}(0) = 0
\end{equation}
in order to glue the solution to $\tau(t) = t$ for $t<0$.
Then, the relation \eqref{eq:bdy eom in evap} for $t > 0$ can be solved as
\begin{equation}
	\label{eq:tau sol}
	\tau(t) = \frac{2}{\alpha A}\left( \frac{I_0(\alpha)K_0(\alpha e^{-\frac{1}{2}At}) - K_0(\alpha)I_0(\alpha e^{-\frac{1}{2}At})}{I_1(\alpha)K_0(\alpha e^{-\frac{1}{2}At}) + K_1(\alpha)I_0(\alpha e^{-\frac{1}{2}At})} \right),
\end{equation}
where $\alpha = \frac{24\pi}{N}\sqrt{\frac{E_0}{\kappa}}$ and $A = \frac{\kappa N}{12\pi}$. Here, $I_\nu$ and $K_\nu$ are modified Bessel functions.
In addition, we assume the Stefan-Boltzmann law as,
\begin{equation}
\label{eq:Stefan-Boltzmann}
	\dv{E}{t}=-\sigma T^2,
\end{equation}
where $\sigma=N(\pi/12)$.
In particular at $t=0$, the boundary stress tensor \eqref{eq:bdy stress tensor unruh} and its derivatives are obtained as
\begin{align}
	E(0)&= E_0 + \frac{N}{12\pi}(b^+ - b^-),\label{eq:bdy stress tensor at t=0}\\
	\dot{E}(0) &=-\sigma T^2(0)= -\frac{N}{12\pi}\left[ (b^+)^2-(b^-)^2 \right],\label{eq:bdy stress tensor dv at t=0}\\
	\ddot{E}(0) &=-2\sigma T(0)\dot{T}(0)= -\frac{N}{6\pi}\left[ \kappa E_0 b^+ - (b^+)^3 + (b^-)^3 \right],\label{eq:bdy stress tensor ddv at t=0}
\end{align}
where we used Eqs.~\eqref{eq:gluing cond} and \eqref{eq:Stefan-Boltzmann}.
From Eq.~\eqref{eq:bdy stress tensor at t=0}, it would be natural to require $b^+ = b^-$
since we throw an infalling pulse with the magnitude $E_0$ at $t=0$, which subsequently indicates $T(0)=0$
in Eq.~\eqref{eq:bdy stress tensor dv at t=0}.
Assuming $\dot{T}(0)$ to be finite in Eq.~\eqref{eq:bdy stress tensor ddv at t=0}, one can get $b^+ = 0$ from the right hand side of \eqref{eq:bdy stress tensor ddv at t=0}.
Hence, the vacuum condition~\eqref{eq:bdy Omega in evap} reduces to $\Omega_u(t) = \frac{1}{2} \log \dot{\tau}(t)$ and $ \Omega_v(t) = 0$,
so that the energy for $t>0$ is obtained as
\begin{equation}
	\label{eq:bdy stress tensor in evap}
	E(t) = E_0e^{-At} \left( \frac{I_1(\alpha) K_2(\alpha e^{-\frac{1}{2}At}) + K_1(\alpha) I_2(\alpha e^{-\frac{1}{2}At})}{I_1(\alpha) K_0(\alpha e^{-\frac{1}{2}At}) + K_1(\alpha) I_0(\alpha e^{-\frac{1}{2}At})} \right)
\end{equation}
while $E(t)=0$ for $t<0$ since $\tau=t$.
In Fig.~\ref{fig:energy}, the black hole energy is depicted by the solid curve,
which shows that it is zero before collapsing
and it is $E_0$ just after collapsing due to the pulse at $t=0$.
Then it decreases and eventually vanishes.

Next, the net flux thrown into the spacetime is obtained by
differentiating the black hole energy~\eqref{eq:bdy stress tensor in evap} with respect to the
boundary time as
\begin{equation}
	\label{eq:diff bdy stress tensor}
	\dv{E}{t} = -\frac{\kappa N}{12\pi}E_0e^{-At}\left( \frac{I_1(\alpha) K_1(\alpha e^{-\frac{1}{2}At}) -K_1(\alpha) I_1(\alpha e^{-\frac{1}{2}At}) }{I_1(\alpha) K_0(\alpha e^{-\frac{1}{2}At}) +K_1(\alpha) I_0(\alpha e^{-\frac{1}{2}At})} \right)^2,
\end{equation}
and the Stefan-Boltzmann law \eqref{eq:Stefan-Boltzmann} leads to the black hole temperature for $t > 0$
\begin{equation}
	\label{eq:temperature}
	T(t) = T_{\rm H}e^{-\frac{1}{2}At}\left( \frac{I_1(\alpha) K_1(\alpha e^{-\frac{1}{2}At}) -K_1(\alpha) I_1(\alpha e^{-\frac{1}{2}At}) }{I_1(\alpha) K_0(\alpha e^{-\frac{1}{2}At}) +K_1(\alpha) I_0(\alpha e^{-\frac{1}{2}At})} \right),
\end{equation}
where $T(0)=0$, $\dot{T}(0) = \kappa E_0$ which is finite, and $T_{\rm H}$ is the Hawking temperature.

As shown in Fig.~\ref{fig:temperature}, our calculations for the black hole temperature is zero for $t<0$ and
the maximum value of it occurs when $t_c =\frac{2}{A} \log\frac{T_{\rm H}}{T_{\rm max}}$,
where $T_{\rm max}=T(t_c)$ and $\frac{d T}{dt}|_{t = t_c} = 0$. The maximum temperature cannot exceed the Hawking temperature in the evaporating black hole.
The black hole temperature starts from zero just when it is formed by the infalling shock wave. It then increases and reaches its maximum temperature at a finite time before gradually decreasing. The temperature profile after $t_c$ is compatible
with the previous result in Ref.~\cite{Engelsoy:2016xyb}.
However, the different solutions to vacuum conditions \eqref{def:evap cond} for the second term in Eq.~\eqref{eq:bdy stress tensor wrt tau} renders the earlier stage of the temperature profile different from the previous one.

\section{discussion}
Using the dynamical boundary equation of motion proposed in Ref.~\cite{Engelsoy:2016xyb}, we obtained the holographic energy and temperature for the exactly soluble dynamical AdS$_2$ black holes by taking into account quantum backreaction. To ensure the validity of the energy expression~\eqref{eq:bdy stress tensor wrt tau}, we first
investigated the black hole in the Hartle-Hawking vacuum state and confirmed that its energy and temperature were compatible with those of the previous work.
In the evaporating black hole of our main interest, we obtained the different type of energy for the black hole in the Unruh vacuum state. Consequently, we found that at the very early stage of evaporation of the black hole, it is decaying slowly compared to the previous result so that the temperature proportional to the decay rate of the black hole energy starts from zero temperature. The temperature increases and reaches a peak at the critical time, and then vanishes eventually. In addition, the peak temperature never exceeds the Hawking temperature during evaporation.

Regarding the temperature behaviour mentioned above, the vanishing initial temperature and the existence of the upper bound may not be the first time. For example, in the RST model~\cite{Russo:1992ax}, the black hole temperature $ T(\sigma^-) = T_{\rm H} [1 - (1 + \frac{M}{\lambda}e^{\lambda(\sigma^- - \sigma_0^+)})^{-2}]^{1/2}$ at future infinity \cite{Liberati:1994za,Eune:2014tma} starts from zero temperature initially and increases according to the monotonically increasing Hawking radiation, eventually vanishing after emission of the thunderpop energy. In this case, the maximum temperature during evaporation does not exceed the Hawking temperature: $ T_{\rm max}(\sigma_{\rm s}^-) = T_{\rm H}[1 - e^{-\frac{96M}{N \lambda}}]^{1/2}$ \cite{Kim:1995wr}, where $\sigma_s^-$ is the conformal time just before the black hole is completely evaporated.

In our work, the temperature of the black hole is assumed to be quasi-static, and thus the thermal bath can be maintained at the same temperature as the black hole.
The black hole temperature vanishes at $t = 0$, which means that the black hole has not yet emitted Hawking radiation to the bath at that instant. If the thermal effect of the black hole originates from pair creation around the black hole, then it may take a finite amount of time to thermalize, similar to the Hayden-Preskill protocol~\cite{Hayden:2007cs}.

A final comment is in order. The Page curve in the evaporating AdS$_2$ black hole~\cite{Almheiri:2019psf} shows that the entanglement entropy for the black hole and radiation system is increasing before the Page time and then decreasing after the Page time. In our temperature calculations, the maximal thermal temperature appears during the evaporation of the black hole at the critical time.
It would be interesting to investigate whether there exists a correlation between the critical time for the maximum temperature and the Page time, in the context of the information loss paradox.

\label{sec:conclusion}

\acknowledgments
We would like to thank Sang-Heon Yi for exciting discussions.
This research was supported by Basic Science Research Program through the National Research Foundation of Korea(NRF) funded by the Ministry of Education through the Center for Quantum Spacetime (CQUeST) of Sogang University (NRF-2020R1A6A1A03047877).
This work was supported by the National Research Foundation of Korea(NRF) grant funded by the Korea government(MSIT) (No. NRF-2022R1A2C1002894).


\bibliographystyle{JHEP}       

\bibliography{reference}

\end{document}